# Depth-Resolved Subsurface Defects in Chemically Etched SrTiO$_3$

Jun Zhang, D. Doutt, and T. Merz

Department of Physics, Ohio State University, Columbus, Ohio 43210

J. Chakhalian, M. Kareev, and J. Liu

Department of Physics, University of Arkansas, Fayetteville, Arkansas 72701

L. J. Brillson

Department of Electrical and Computer Engineering and Department of Physics,

Ohio State University, Columbus, Ohio 43210

Depth-resolved cathodoluminescence spectroscopy of atomically flat TiO$_2$-terminated SrTiO$_3$ single crystal surfaces reveals dramatic differences in native point defects produced by conventional etching with buffered HF (BHF) and an alternative procedure using HCl-HNO$_3$ acidic solution (HCLNO), which produces three times fewer oxygen vacancies before and nearly an order of magnitude fewer after pure oxygen annealing. BHF-produced defect densities extend hundreds of nm below the surface, whereas the lower HCLNO-treated densities extend less than 50 nm. This "Arkansas" HCLNO etch and anneal method avoids HF handling and provides high quality SrTiO$_3$ surfaces with low native defect density for complex oxide heterostructure growth.

Transition metal perovskite oxides are materials with a broad spectrum of functional properties. Of particular interest are perovskite oxide heterostructures with atomically flat interfaces that exhibit functionality not present in their constituent oxides, [1-5] e.g., highly mobile electron systems induced at the $LaAlO_3/SrTiO_3$ interface.[1] One of the critical factors that can strongly affect the electronic properties of complex oxide heterointerfaces is the oxygen-related defect.[6-8] Chemical treatments used to obtain atomically-flat substrates for epitaxial perovskite growth may create surface and subsurface oxygen vacancy defects, introducing localized states that can strongly impact the heterostructures' electronic properties.[2, 3, 8, 9]

Crystalline $SrTiO_3$ is the most commonly used substrate for perovskite growth. The conventional procedure to prepare atomically flat $TiO_2$-terminated $SrTiO_3$ surfaces employs a buffered HF (BHF) treatment.[10] Recently, an alternative "Arkansas" etching method using a 3:1 $HCl-HNO_3$ (HCLNO) solution[11] has been reported to produce high-quality $TiO_2$-terminated surfaces with equivalent morphology. Photoluminescence spectroscopy and soft x-ray absorption spectroscopy showed that oxygen vacancy defect states are significantly reduced in the HCLNO treatment compared with the BHF treatment.[11] However, the spatial distribution and magnitude of defects created with these two treatments are still not clear, especially at the free surface where growth of complex oxide commences. In this Letter we report a comparative study of surface and near-surface defects of $SrTiO_3$ single crystals treated by these two methods by depth-resolved cathodoluminescence spectroscopy (DRCLS). Here, the excitation depth of incident electron beams vary with energy, so DRCLS can generate near band edge (NBE) and defect optical transitions at surfaces, interfaces and the bulk, providing unique nanoscale-resolved information about defect distributions.[12-14] Our results show that the HCLNO treatment creates much less oxygen vacancies at all measured depths compared with the BHF treatment, and the density of oxygen vacancies decreases away from the surface, indicating that these residual defects locate mostly in the near-surface region.

$TiO_2$-terminated $SrTiO_3$(100) single crystal substrates with atomically flat surfaces were obtained by BHF and HCLNO treatments, including chemical etching and subsequent annealing in air at 1000 °C for 30 minutes.[11] DRCLS measurements were performed in an ultra high vacuum chamber at ~ 42 K. The constant 2 μA incident electron beam focused to 0.25 cm$^2$, beam voltage $E_B$ varied from 1 kV to 5 kV and spectra acquisition time of 5 seconds corresponds to fluences more than two orders of magnitude below threshold fluences required to observe 1 keV electron beam effects on $TiO_2$.[15] Indeed, such effects must be small relative to the dramatic differences reported here between BHF- and HCLNO-etched surfaces, as well as their decreases with surface processing and depth distribution – all obtained under the same excitation conditions. For $E_B$= 1, 2, 3, 4, and 5 keV, Monte Carlo simulations of the electron cascade yield maximum excitation depths of ~ 20, 45, 85, 140, and 210 nm, respectively, below the bare $SrTiO_3$ surface.

The "as-received" $SrTiO_3$ CL spectra in Fig. 1(a) show multiple emissions, and the deconvolved spectra exhibit an NBE emission at $E_G$ = 3.2 eV plus sub-band gap emission peaks centered at 2.1 and 2.7 eV, with examples shown for the 20 nm depth. The 3.2 eV peak can be considered as the emission in the non-relaxed sample, i.e., without excitonic effects, while the emissions at lower energies of 2.1 eV and 2.7 eV are from charge-transfer-vibronic excitons pinned by oxygen vacancies ($V_O$) and surface.[16-18] The weak 1.6 eV emission can be attributed to defects having a $Ti^{3+}$ state surrounded by small numbers of oxygen ions,[19-21] such as, for instance, the $Ti^{3+}$ states at the crystallographic step on the surface of $TiO_2$.[11]

Both BHF and HCLNO treatments produce strong changes in near-surface defect emissions. Similar to the as-received crystals, the deconvolved CL spectra of chemically-etched $SrTiO_3$ show NBE emission at 3.2 eV and $V_O$-related emissions at 2.1±0.1 eV and 2.6±0.1 eV, as shown in Fig. 1(b) and Fig. 2(a), but the defect emissions intensities differ significantly and depend on the etching solution used. After BHF treatment, i.e., acid etching and post-annealing in air, $V_O$-related peaks at 2.1 eV and 2.5 eV of the deconvolved spectra in Fig. 1(b) are an order of magnitude more intense than those shown in Fig.1(a) for as-received $SrTiO_3$. Figure 3 plots the intensity ratios of the 2.1 eV oxygen vacancy peak to 3.2 eV NBE emission, I(2.1 eV)/I(3.2 eV), for the as-received and processed $SrTiO_3$. Clearly the conventional BHF treatment increases oxygen vacancies dramatically not only at the surface but also extensively deeper than 200 nm.

HCLNO-treated $SrTiO_3$ showed much weaker defect emissions. Figure 3 shows that $V_O$-related features only double compared with as-received crystals, primarily within 20 nm of the surface. I(2.1 eV)/I(3.2 eV) for HCLNO-treated crystals increases vs. as-received $SrTiO_3$ but remains much smaller than for BHF-treated crystals. These results clearly imply that the HCLNO treatment creates far fewer oxygen vacancies and electronically-active sites localized near the surface compared to the BHF treatment. In addition, Fig. 2(a) shows that the 1.6 eV peak is more intense at the HCLNO-treated crystal surface, in line with the existence of the crystallographic steps on the HCLNO-treated surface.[11]

To further reduce defect densities, we annealed the HCLNO-treated $SrTiO_3$ in oxygen at 650 °C for 30 minutes. Figure 2(b) shows a dramatic reduction of the 1.6 eV $Ti^{3+}$ feature in the HCLNO-treated crystal after pure oxygen annealing. Furthermore, Fig. 3 shows that the intensity ratio I(2.1 eV)/I(3.2 eV) decreases at all depths after annealing, becoming comparable to that for the as-received crystal. This is consistent with the fact that oxygen annealing partially fills oxygen vacancies and converts some $Ti^{3+}$ to $Ti^{4+}$, reducing both types of defects, notwithstanding their apparently independent behavior. Therefore, combined with oxygen annealing, the alternative HCLNO treatment can create very high-quality $SrTiO_3$ surfaces that have not only morphology comparable to the conventional BHF method but also low densities of electronically-active defects that, more importantly, are limited overall to within the first few tens of nanometers of the surface. Besides lower defect surfaces, a practical advantage of the HCLNO method is the avoidance of HF which introduces additional safety and disposal issues.

We used a Kelvin probe force microscope to map electrostatic potential variations near asperities in these surfaces and gauge the potential screening by free carriers away from charge sites. The estimated average excess charge due to defects at the $SrTiO_3$ surfaces[22] is ~ 1.6 x$10^{15}$ $cm^{-3}$ for the HCLNO-treated and annealed surface and 1.4 x $10^{16}$ $cm^{-3}$ for the BHF-treated surface, consistent with the difference in oxygen vacancy to NBE intensity ratios in Fig. 3 for these surfaces.[23]

The pronounced differences between etching processes may be due to the very oxidizing nature of the HCl and $HNO_3$ combination because of the $NO_3^-$ groups versus the less oxidizing BHF etch. With HCLNO's oxidizing nature, there is little reduction of the $Ti^{3+}$ and oxygen vacancies must be compensated by cation vacancies. Further, HF is rather reactive with $SrTiO_3$, for example, converting $TiO_2$ crystallites to $TiOF_2$ so that the fluoride ions likely react with Ti atoms in $SrTiO_3$ rather aggressively. Such extended effects are consistent with dislocations in $SrTiO_3$ that can extend as deep as 10 μm into the bulk[24] so that defects created at the surface can diffuse hundreds of nanometers below after high-temperature annealing.[25]

In summary, DRCLS measurements demonstrate that the conventional BHF treatment creates high densities of oxygen vacancies in $SrTiO_3$ crystals, while the alternative HCLNO treatment creates far fewer oxygen vacancies that can be further reduced by oxygen annealing. Moreover, these oxygen vacancies in the HCLNO-treated $SrTiO_3$ are located primarily within tens of nm of the surface rather than forming a reservoir of defects extending hundreds of nm into the bulk that could diffuse under elevated growth temperatures. Hence, HCLNO etch combined with high temperature oxygen annealing is a very promising procedure to create atomically-smooth $SrTiO_3$ surfaces without introducing high native point defect densities, a prerequisite for controlling electronic structure at perovskite heterointerfaces.

OSU authors gratefully acknowledge support from ONR Grant N000140-08-1-1090 and NSF Grant DMR-0820414. University of Arkansas authors acknowledge support from DOD-ARO under Contract No.0402-17291 and NSF Contract No. DMR-0747808.

Authors acknowledge useful discussions with OSU Prof. P. Woodward and U. Arkansas Prof. S. Prosandeev.

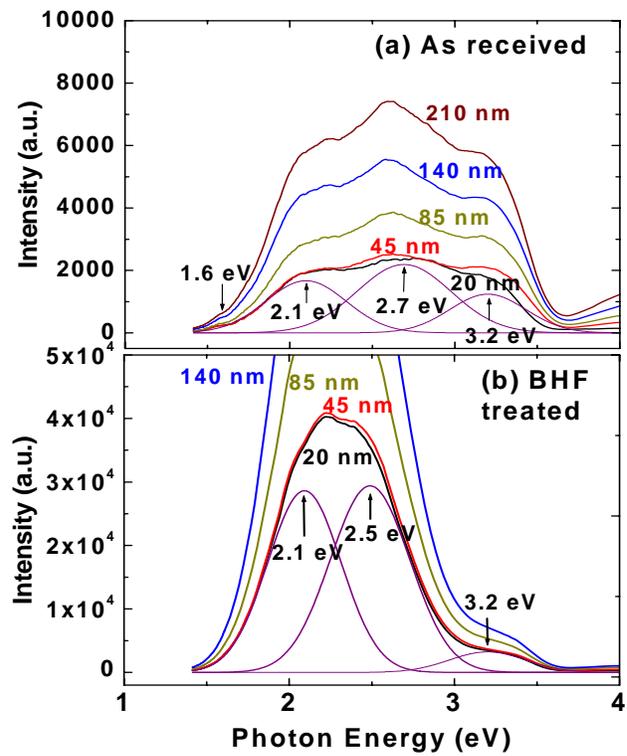

Fig. 1. Representative CL spectra of an (a) as-received vs. (b) BHF-treated SrTiO$_3$ single crystal. Deconvolved peaks (purple, denoted by arrows) are shown only for 20 nm depth as examples.

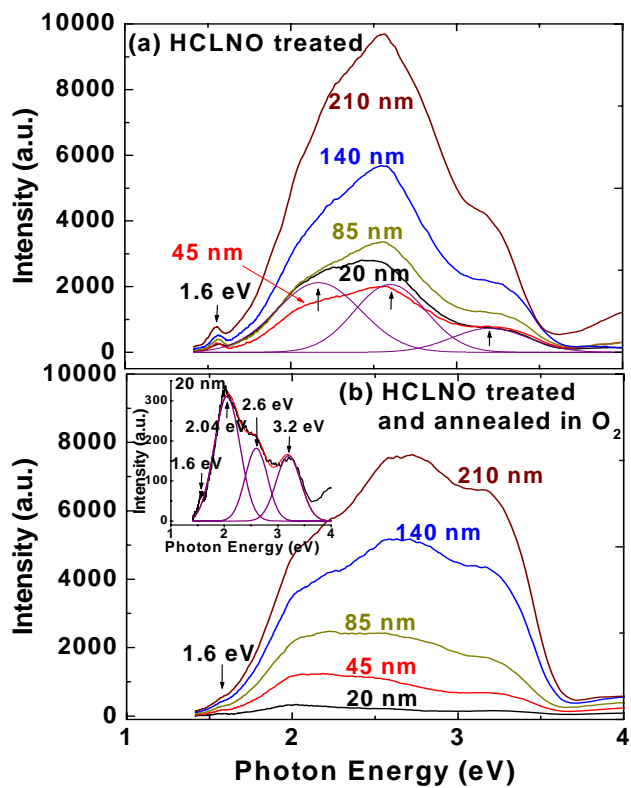

Fig. 2. Representative CL spectra of a HCLNO-treated $SrTiO_3$ single crystal (a) before and (b) after high temperature oxygen anneal. Deconvolved peaks (purple, denoted by arrows) are shown only for 20 nm depth as examples in (a) and the inset of (b).

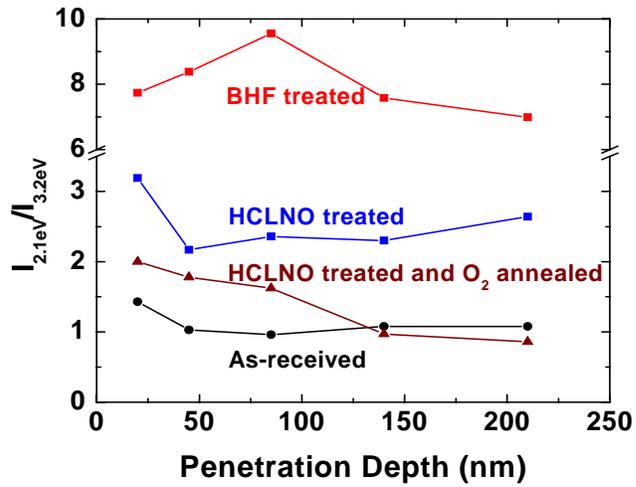

Fig. 3. Depth dependence of intensity ratios of the deconvolved oxygen vacancy (2.1 eV) to NBE (3.2 eV) emission features in the CL spectra of as-received and chemically-etched $SrTiO_3$ single crystals.